\documentclass[12pt,leqno]{article}
\usepackage{latexsym}
\usepackage{amsfonts}
\usepackage{amsmath}
\usepackage{amsthm}

\def\d{\partial}

\renewcommand{\exp}[1]{{\rm exp}(#1)}

\newcommand{\qcommut}[2]{{[#1\stackrel{*}{,}#2]}}
\newcommand{\commut}[2]{[#1,#2]}

\newcommand{\tr}{{\rm \,Tr}}
\def\half{{\frac{1}{2}}}

\newcommand{\mysection}[1]{\section{#1}\setcounter{equation}{0}}
\def\be{\begin{eqnarray}}
\def\ee{\end{eqnarray}}
\def\beann{\begin{eqnarray*}}
\def\eeann{\end{eqnarray*}}
\def\beq{\begin{equation}}
\def\eeq{\end{equation}}
\def\ba{\begin{array}}
\def\ea{\end{array}}
\def\ben{\begin{enumerate}}
\def\een{\end{enumerate}}
\def\bea{\begin{eqnarray}}
\def\eea{\end{eqnarray}}
\def\beann{\begin{eqnarray*}}
\def\eeann{\end{eqnarray*}}
\def\beq{\begin{equation}}
\def\eeq{\end{equation}}
\def\ba{\begin{array}}
\def\ea{\end{array}}
\def\ben{\begin{enumerate}}
\def\een{\end{enumerate}}

\def\5{\bar }
\def\6{\partial }
\def\7{\hat }
\def\4{\tilde }

\def\s0#1#2{\mbox{\small{$\frac{#1}{#2}$}}}

\def\qed{\hbox{${\vcenter{\vbox{
\hrule height 0.4pt\hbox{\vrule width 0.4pt height 6pt
\kern5pt\vrule width 0.4pt}\hrule height 0.4pt}}}$}}

\begin{document}

\begin{titlepage}
\begin{flushright}
ULB-TH/01-17\\
hep-th/0106188
\end{flushright}

\begin{centering}

\vspace{0.5cm}

\huge{Seiberg-Witten maps from the point of view of consistent deformations
of gauge theories}\\ 

\vspace{.5cm}

\large{Glenn Barnich$^{a,*}$, Maxim Grigoriev$^{b}$ and 
Marc Henneaux$^{a,c}$
}

\vspace{.5cm}

$^a$
Physique Th\'eorique et Math\'ematique,\\ Universit\'e Libre de
Bruxelles,\\
Campus Plaine C.P. 231, B--1050 Bruxelles, Belgium

\vspace{.5cm}

$^b$ Lebedev Physics Institute,\\
53 Leninisky Prospect,\\
Moscow 117924, Russia

\vspace{.5cm}

$^c$ Centro de Estudios Cientif\'{\i}cos, \\
Casilla 1469, Valdivia, Chile

\vspace{.5cm}

\end{centering}

\begin{abstract}
Noncommutative versions of theories with a gauge freedom define
(when they exist) consistent deformations of their commutative
counterparts.  
General aspects of Seiberg-Witten maps are discussed from this
point of view.  In particular, the existence of the
Seiberg-Witten maps for various noncommutative theories is
related to known cohomological theorems on 
the rigidity of the gauge symmetries of
the commutative versions.  In technical terms,
the Seiberg-Witten maps
define canonical transformations in the antibracket
that make the solutions of the master equation
for the commutative and noncommutative versions
coincide in their antifield-dependent terms.
As an
illustration, the 
on-shell
reducible noncommutative Freed\-man-Town\-send theory
is considered. 
\end{abstract}

\vspace{.5cm}

\footnotesize{$^*$ Research Associate of the Belgium National Fund for
  Scientific Research}

\vfill

\end{titlepage}

\mysection{Introduction}

A remarkable feature of Yang-Mills theory is the 
(formal) rigidity of its gauge
structure.  Namely, there is no consistent deformation
of the Yang-Mills action
\begin{equation}
I_{YM}[A_\mu] = - \frac{1}{4} \int d^n x\ F^a_{\mu \nu}
F^{\mu \nu}_a
\label{commYMaction}
\end{equation}
that truly deforms its gauge symmetries.  By allowed redefinitions
of the gauge parameters and of the fields, 
one can always bring the gauge transformations
of the deformed theory back to the original form
\begin{equation}
\delta_\epsilon A^a_\mu = \partial_\mu \epsilon^a + f^a_{\; \; b c}
A^b _\mu \epsilon^c .
\end{equation}
In this light, the existence of the so-called Seiberg-Witten
(SW) map \cite{Seiberg:1999vs}
for the noncommutative deformation of (1.1) has a
clear underlying algebraic rationale.
[For a recent
review on noncommutative Yang-Mills theories, see \cite{Douglas:2001ba}.]

The rigidity of the gauge symmetries of (\ref{commYMaction}) was
established in \cite{Barnich:1994ve,Barnich:1995mt} by
cohomological techniques, without assuming Lorentz invariance
or restricting the class of possible
deformations to those with pre-assigned ``engineering 
dimension".
This is particularly relevant to the
case where a prescribed (``external")
constant matrix $\theta^{\mu\nu}$ with (negative) mass dimension is
present\footnote{ Actually, rigidity holds only if the gauge group
has no Abelian factor. When there is an Abelian factor, there can
be deformations of (\ref{commYMaction}) that truly deform the
gauge structure; these involve conserved currents
\cite{Barnich:1994ve}. One can easily show that such a possibility
does not arise in the noncommutative deformation of $U(N)$
Yang-Mills theory because of the derivative structure of the
coupling, see  \cite{Barnich:2001bb}.}.

In this paper, we discuss
Seiberg-Witten maps from the point of view of consistent deformations
of gauge theories in the context of the Batalin-Vilkovisky formalism
for local gauge theories
\cite{Zinn-Justin:1974mc,Zinn-Justin:1989mi,deWit:1978cd,
Batalin:1981jr,Batalin:1983wj,Batalin:1983jr,Batalin:1984ss,
Batalin:1985qj,Fisch:1990rp,Henneaux:1991rx} (for reviews, see 
e.g. \cite{Henneaux:1992ig,Gomis:1995he}). In particular, we derive
cohomological conditions that guarantee the existence of Seiberg-Witten
maps.  The SW maps turn out to be in fact canonical transformations
in the antibracket that generically
mix the original fields, the ghosts and the antifields.  The 
antifields do not occur in the transformation of the
fields and ghosts for the Yang-Mills case because the gauge
structure is defined then not just on-shell but also off-shell.  However,
they do occur for more general gauge theories.

We illustrate this feature for the noncommutative deformation
of the Freed\-man-Town\-send model \cite{Freedman:1981us},
whose gauge structure is rigid
(see \cite{Henneaux:1996ts} and section \ref{commft}). This deformation 
exhibits clearly the new features of
generic gauge theories which are: (i) the gauge
symmetries are reducible, so ghosts of ghosts are necessary 
and must be considered in the SW canonical transformations (corresponding
to possible redefinitions of the reducibility coefficients);
(ii) reducibility holds only
on-shell and so, one cannot separate the symmetries from the
dynamics, as one can in the Yang-Mills case. 
We explicitly construct the generating functional
of the SW map in the $u(1)$-case
and point out the mixture of the antifields with
the fields, which precisely reflects the mixing
of the dynamics with the symmetries.

Our conclusions are presented in section \ref{5}.

Finally, we mention a few references where other approaches to 
SW maps in the Yang-Mills case are discussed. Existence by 
explicit construction has been shown in e.g.\cite{Okuyama:1999ig},
where commutative and  noncommutative versions of Wilson lines were
compared and in \cite{Jurco:2000fs}, where the SW map was computed in
the framework of Kontsevich's approach to deformation quantization. A
reference that also uses cohomological arguments (though not in the BV
formalism) is \cite{Brace:2001fj}. An explicit inverse SW map was
given in \cite{Okawa:2001mv}.

\mysection{Seiberg-Witten maps: general considerations\label{s3}}

\subsection{Noncommutative gauge theories as consistent deformations of
  commutative gauge theories}

Consider a gauge theory with action
${I_0}[\hat\varphi] = \int d^n x\ L_0([\hat \varphi])$,
where the notation $f = f([y])$ means that the function
$f$ depends on the variable $y$ and a finite number of its derivatives.  
We put a hat on the fields in anticipation of changes of field
variables performed below to unhatted fields.
We denote the infinitesimal gauge
transformations by 
$\delta_{0,\epsilon}\hat\varphi^i ={R_0^{i}}([\hat\varphi],[\hat
\epsilon])$  (the
dependence on the gauge parameters $\hat
\epsilon^\alpha$ and their derivatives
is of course linear).  One form of the Noether identities
expressing gauge invariance is
\bea 
\frac{\delta
  {L_0}}{\delta\hat\varphi^i}{R_0^{i}}
+\partial_\mu j^\mu=0, 
\eea 
for some $j^\mu$, where $\delta L_0 /\delta\hat\varphi^i$ are the
Euler-Lagrange derivatives of $L_0$.  The gauge theory may be reducible,
i.e., there may exist particular choices of the gauge
parameters, $\hat
\epsilon^\alpha={Z_0}^{\alpha}([\varphi], [\hat \eta])$ for
which the gauge transformations are trivial,
\bea
{R_0^{i}}([\varphi], [{Z_0}]) \approx 0,
\eea 
where $\approx$ means on-shell for the equations of motion
defined by ${I_0}$.  As we have indicated, there may be
more than one reducibility condition, i.e., the ${Z_0}^{\alpha}$'s
may depend (linearly) on a certain number of reducibility
parameters $\hat \eta^{A}$ and their derivatives.

A (formal) consistent deformation of such a
gauge theory is defined by giving (i) a deformed action,
(ii) deformed gauge symmetries and (iii) deformed reducibility 
expressions
\bea
\hat I={I_0}+g{I_1}+\frac{1}{2}g^2{I_2}
+\dots,\\
\hat {R^i}={R_0^{i}}+g{R_1^{i}}
+\frac{1}{2}g^2{R_2^{i}}+\dots,\\
\hat Z^\alpha={Z_0}^{\alpha}
+g{Z_1}^{\alpha}+\frac{1}{2}g^2 {Z_2}^{\alpha}+\dots,
\eea such that the Noether identities and the reducibility
equations are preserved order by order in the deformation
parameter $g$,
\bea \frac{\delta                                     
  {\hat L}}{\delta\hat\varphi^i}\hat {R^i} 
+\partial_\mu \hat j^\mu=0, \\
\hat {R^i} ([\varphi], [\hat {Z}])\approx^\prime
0, \eea 
where $\approx^\prime 0$ means on-shell for the equations
of motions defined by the complete action $\hat I =
\int d^n x\ \hat L$.  The deformed Lagrangian $\hat L$ may
involve all the derivatives of the fields, but we require that
each term $L_i$ in the power expansion $\hat L = L_0 + g L_1 + \cdots$
be a local function, i.e., contains a finite number of derivatives.

With these definitions, noncommutative extensions of
commutative gauge theories 
are clearly
(when they exist)  consistent deformations of their
commutative counterparts,
the deformation parameter $g$ multiplying the
matrix $\theta^{\mu\nu}$ defining the non-commutativity of
the coordinates (see below).
                                      
\subsection{Gauge structure}

As explained e.g. in \cite{Henneaux:1992ig}, the most general 
gauge transformation is obtained by adding to the transformations
$\delta_{0,\epsilon} \hat \varphi^i = R^i_0$ (or
$R^i$), in which the gauge parameters are chosen arbitrarily
(in particular, can be functions of the fields and their
derivatives), an arbitrary antisymmetric combination
of the equations of motion.
Because the most general
gauge transformation explicitly refers to the dynamics, the
algebra of all the gauge transformations of pure Maxwell
theory, say, is different from the algebra of Born-Infeld
theory.  However, in both cases, the relevant information
about the gauge transformations
is contained in the transformation $\delta A_\mu =
\partial_\mu \epsilon$, which is identical for the
two theories. 
Only the ``on-shell trivial" gauge transformations,
involving the equations of motion, are different.  For this
reason, one says that pure Maxwell theory and
Born-Infeld theory have identical gauge structures.

More generally, one says that two gauge theories
have identical gauge structures if it is possible
to choose their generating sets such that they coincide.
The generating sets are precisely the subsets of the gauge
algebra that contain the relevant information
about all the gauge symmetries, in the sense that any
gauge transformation can be obtained from the generating
set by choosing appropriately the gauge
parameters (possibly, as functions of the fields) and
adding if necessary an on-shell trivial gauge symmetry
\cite{Henneaux:1992ig}.  There is a huge freedom in
the choice of generating sets.  Furthermore, one may
also redefine the reducibility relations\footnote{Note that if
reducibility holds only on-shell, the equations of
motion that appear in the reducibility identities must
be the same in both theories - when written in the
same variables.}.  Moreover, the
equivalence may become manifest only after one has redefined the
field variables. So in order to show that two theories have identical
gauge structures, one must establish the existence of a field
transformation such that the two generating sets can be made to
coincide (through redefinitions of the gauge parameters and addition
of on-shell trivial gauge transformations).

One virtue of the antifield formalism is that
all this freedom can be neatly taken into account
trough canonical transformations 
(see e.g. \cite{Voronov:1982cp,Batalin:1984ss,Henneaux:1992ig}).
For this reason, we review how consistent deformations
are formulated in the antifield (BV) formalism.

\subsection{Cohomological reformulation of consistent deformations in the
BV formalism\label{s3.2}}

In the framework of the BV formalism, all the information on the
invariance of the action and the algebra of gauge symmetries
is encoded in an extended action satisfying the so-called master equation.
The problem of consistent deformations of gauge theories
can then be reformulated
\cite{Barnich:1993vg} (see \cite{Henneaux:1997bm} for a review) as the
problem of deforming the solution of the master equation in the space
of local functionals (while maintaining the
master equation itself). In the present context, local functions
are formal power series in the deformation parameter, each term
depending on the original fields, the ghosts, ghosts for
ghosts, their antifields and a finite number of derivatives of all
these fields. Local functionals are identified with equivalence
classes of local functions up to total divergences
(see\cite{Barnich:2000zw} for more details).
We shall denote the solution of the (classical) master equation for the
original and deformed theories by $S_0$ and $\hat S$, respectively.
One has
\bea 
S_0 &=&I_0+\int d^nx\ \hat
\varphi^*_i{R_0^{i}}([\hat \varphi],[\hat C])+
\hat C^*_\alpha{Z_0}^{\alpha}([\hat \varphi],[\hat \rho])+\dots,\\
\frac{1}{2} \! && \! \! \! \! \! \! 
\! \! \! (S_0,S_0)=0,\\
\hat S &=& S_0 + g S_1 +(1/2)g^2 S_2+\dots \nonumber\\
&=&\hat I+\int d^nx\ \hat \varphi^*_i \hat {R^{i}}([\hat \varphi],
[\hat
C])+
\hat C^*_\alpha\hat {Z}^{\alpha}([\hat \varphi],[\hat \rho])+\dots,\\
\frac{1}{2} \!&& \! \! \! \! \! \!
\! \! \! (\hat S,\hat S)= 0.\label{def}
\eea
Here, the $\hat C^\alpha$ are the ghosts replacing the gauge
parameters, the $\hat\rho^A$ are the ghosts for ghosts, while
$\hat\varphi^*_i$, $\hat C^*_\alpha$ and $\hat\rho^*_A$ are the
associated antifields, of respective antifield number $1,2,3$
(see
e.g. \cite{Batalin:1981jr,Henneaux:1992ig,Gomis:1995he}
for details).

Note that the antifield-independent part of the solution
of the master equation is just the classical action.  The
information about the gauge symmetries, their algebra, the reducibility
equations etc is contained in the antifield-dependent part.  {\em Thus,
if the original and deformed theories have the
same gauge structure, $S_0$ and $\hat S$  have the
same antifield-dependent part and vice-versa.}
The advantage of the cohomological reformulation is that standard
techniques of deformation theory can now be applied.                      

In particular, (\ref{def}) implies that
\bea
(\hat S,\frac{\partial \hat S}{\partial g})=0,\label{coc}
\eea
which in turn implies that an infinitesimal deformation $
\frac{\partial \hat S}{\partial g}|_{g=0}=S_1$ is a cocycle of the
BRST differential $s_0=(S_0,\cdot)$ of the undeformed theory,
\bea
(S_0,S_1)=0.
\eea

A deformation is
trivial if it can be undone through a canonical, i.e., antibracket
preserving, transformation\footnote{ 
Only canonical transformations that
reduce to the identity to order $0$ in the deformation parameter are
considered here. Invertibility of these transformations in the space
of formal power series is then guaranteed.}:
\bea
\hat S[\hat\phi([\phi],[\phi^*];g),\hat\phi^*([\phi],[\phi^*];g);g]=
S_0[\phi,\phi^*],\label{triv0}
\eea
Here  -- and throughout below --,
  the original fields and ghosts are collectively denoted by
  $\hat \phi$, while $\hat \phi^*$ denotes collectively all the antifields.
Thus, a generic canonical transformation
mixes the original fields, the ghosts and the antifields.
Equivalently, this means that $\hat S$ can be obtained from
the original $S_0$ 
through the inverse canonical 
transformation
\bea
\hat S[\hat\phi, \hat\phi^*;g] = S_0[\phi([\hat \phi],[\hat\phi^*];g),
\phi^*([\hat \phi],[\hat \phi^*];g)].
\label{triv1}
\eea
This is the case iff the cocycle $\frac{\partial \hat S}{\partial g}$
is a coboundary of the deformed theory,
\bea
\frac{\partial \hat S}{\partial g}=(\hat S,\hat \Xi).\label{triv}
\eea
Indeed, if (\ref{triv}) holds, then a
canonical transformation
with the required properties
is given by
\begin{equation}
\phi^A(x)=P\exp{\int_0^gdg^\prime
\Big(\cdot,\hat\Xi(g^\prime)\Big)}\hat\phi^A(x), \; \;
\phi^*_A(x)=P\exp{\int_0^gdg^\prime \Big(\cdot,\hat
\Xi(g^\prime)\Big)}\hat\phi^A(x). \label{3.311} 
\end{equation}
Conversely, if the deformed action can be obtained from
the undeformed one by an canonical transformation
for any $g$, then the passage from
$\hat S(g)$ to $\hat S (g + \delta g)$ is an infinitesimal
canonical transformation (by the group property of
canonical transformations) and (\ref{triv}) holds.
It will be useful in the sequel to introduce the
generating functional 
$F[\phi,\hat\phi^*;g]$ of ``second type'' in ghost number $-1$
such that
\bea
\hat\phi^A(x)=\frac{\delta^L
F}{\delta\hat\phi^*_A(x)},\ \phi^*_A(x) =\frac{\delta^L
F}{\delta\hat\phi^A(x)},\label{transfo}
\eea 
(see \cite{Batalin:1984ss} and appendix A of \cite{Troost:1990cu}
for material on antibracket preserving transformations).

It follows from (\ref{coc}) and (\ref{triv}) that a necessary
condition for the existence of a non trivial deformation is the
existence of a non trivial cohomology class for the deformed
theory. Because every cocycle of the BRST differential $\hat s=(\hat
S,\cdot)$  of the deformed theory gives to
lowest order in $g$ a cocycle of the BRST differential
$s_0=(S_{0},\cdot)$ of the undeformed one, and furthermore, every
coboundary of the undeformed theory can be extended to a
coboundary of the deformed theory, it follows that
non trivial
deformations are controlled by the local BRST cohomology of the
undeformed theory. In particular, if this cohomology is empty in
the relevant subspace of the space of local functionals, it
follows that the deformation is trivial. By relevant subspace, we
mean the subspace of local functionals of ghost number $0$ 
possibly restricted through
additional requirements like global symmetries or power counting
restrictions, depending on the problem at hand.

Elements of $H^0(s_0)$ are called non trivial infinitesimal
deformations. One can furthermore show that the obstruction to
extending infinitesimal deformations are controlled by the
antibracket map \bea (\cdot,\cdot)_M: H^{0}(s^{(0)})\otimes
H^{0}(s^{(0)})\longrightarrow H^{1}(s^{(0)}) \eea
but this will not be needed here.

\subsection{SW maps in the context of
  deformation theory\label{3.3}}

Given a non trivial consistent deformation of a gauge theory,
one may ask whether the gauge
symmetries, their algebra and their reducibilities
are equivalent to the one of the undeformed theory through allowed
redefinitions of the most general type.  The
allowed redefinitions of the gauge symmetries can involve
redefinitions of the gauge parameters that contain the fields
themselves, as well as the possible addition of
on-shell trivial symmetries \cite{Henneaux:1992ig}.
Note that if the classical action $\hat I$ is equivalent through field
redefinitions to the action $I_{0}$, then the gauge symmetries and
their structure are of course equivalent, whereas the converse does not
hold (e.g., as explained above,
Maxwell theory and Born-Infeld theory are based
on inequivalent actions but their gauge structures are identical).

Because the gauge symmetries and their structure are described in the
master equation through terms with strictly positive antifield number,
this question amounts to the question of the existence of a
canonical transformation
$\hat\phi[\phi,\phi^*;g],\hat\phi^*[\phi,\phi^*;g]$
such that
\begin{equation}
\hat S[\hat\phi[\phi,\phi^*;g],\hat\phi^*[\phi,\phi^*;g];g]=
S_0[\phi,\phi^*] +  V [\varphi, g]
\label{3.35}
\end{equation}
That is, if one can absorb all the antifield-depen\-dence 
through a canonical transformation, the only
effect of the deformation will be indeed (after redefinitions) just to
change the original action $I_0[\varphi]$ into
$I^{\rm eff}[\varphi; g] \equiv I_0[\varphi] +  V [\varphi, g]$, 
\begin{eqnarray}
I_0[\varphi] &\rightarrow& I^{\rm eff}[\varphi; g] =
 I_0[\varphi] +  V [\varphi; g], \\
V [\varphi; g] &=& g V_1[\varphi] + \frac{g^2}{2} V_2[\varphi]
+ \cdots
\end{eqnarray}
without modifying the gauge
transformations.
The fact that one considers general canonical
transformations automatically incorporates all the available freedom
(see e.g. \cite{Henneaux:1992ig}).
In particular, it allows for redefinitions of the ghosts that
involve the original fields, or, what is the same in
terms of the gauge transformations,  redefinitions
of the gauge parameters that contain the fields
$\varphi^i$.   It also allows for redefinitions of the gauge
transformations that involve on-shell trivial gauge symmetries.
In short, it enables one to go from one generating set to any other
generating set.

By a reasoning similar to that of the previous paragraph, the
condition (\ref{3.35}) 
can be shown to be equivalent to the condition
\bea
\frac{\partial \hat S}{\partial g}
=\frac{\partial V}
{\partial g}+(\hat S,\hat \Xi).\label{3.37}
\eea
Note also that Eq. (\ref{3.35})
reads, in
the variables $\phi,\hat\phi^*$, 
\bea
\hat S[\frac{\delta^L F}{\delta\hat\phi^*},\hat\phi^*;g]=          
S_0 [\phi,\frac{\delta^L F}{\delta \phi}] +
V [\varphi; g].
\label{3.38}
\eea

The canonical transformations that achieve (\ref{3.35})
or (\ref{3.38}) will be called Seiberg-Witten (SW) maps.

\subsection{Existence of Seiberg-Witten maps}
An important instance where (\ref{3.37}) holds arises
in the following situation.
To lowest order in $g$, condition (\ref{3.37}) reduces
to \bea S_1= V_1(\hat \varphi) + (S_0, \Xi_1),
\label{3.40} \eea for the
$s_{0}$ cocycle $\hat S_1$, with
$\Xi_{1}=\hat\Xi[\hat\phi,\hat\phi^*,0]$.  Suppose
that in the relevant subspace in which the
deformation is allowed to take place, the representatives of all the
cohomology of $s_{0}$ can be chosen to be antifield independent,
i.e., \bea (S_{0},C)=0\Longrightarrow
C=C^\prime[\varphi]
+(S_{0},D)\label{ass} \eea for some $C^\prime[\varphi]$
that depends only on the original fields and not
on the antifields (and which is of
course annihilated by $s_0$, $s_0 C' = 0$).
Then (\ref{3.35}) can be
achieved through a succession of canonical transformations.

Indeed, let $\hat z=(\hat\phi,\hat\phi^*)$ and consider the
canonical transformation $z^1=\exp g(\cdot,\Xi^{(1)})\hat z$, so
that $\hat z=z^1-g(z^1,\Xi^{(1)})+O(g^2)$. It follows that \bea
\hat S[\hat z([z^1];g);g]=S_{0}[z^1]+g  V_1 
[\varphi^1]+g^2 \tilde S^{(2)}[z^1]+O(g^3), \eea for some
$\tilde S^{(2)}[z^1]$ and $ s_0  V_1 = 0$.
More generally,
assume that the Seiberg-Witten map has been constructed to order $k$,
i.e., that we have constructed a canonical transformation
$z^k=z^k[\hat z;g]$ such that
\bea
\hat S[\hat z[z^k;g];g]=S_{0}[z^k]+\sum_{i=1}^k g^i V_i
[\varphi^k]+g^{k+1} \tilde S^{(k+1)}[z^k]+O(g^{k+2}),
\eea
with $s_0 V_i = 0$.
The equation $\half (\hat S[\hat
z[z^k;g]],\hat S[\hat z[z^k;g]])=0$, which holds because the
transformation is canonical then gives to lowest non trivial order
$g^{k+1}$ that $(S_{0}[z^k],\tilde
S^{(k+1)}[z^k])=0$, which in turn implies by the assumption (\ref{ass})
that
\bea
\tilde S^{(k+1)}[z^k]= V_{k+1}[\varphi^k]+(S_{0}[z^k],
\Xi_{k+1}[z^k]).
\eea
The next canonical transformation is then
$z^{k+1}=\exp g^{k+1}(\cdot,\Xi_{k+1}[z^k])z^k$ so that
$z^k=z^{k+1} - g^{k+1}(z^{k+1},\Xi_{k+1})+O(g^{k+2})$. If $\hat
z[z^{k+1};g]=\hat z[z^{k}[z^{k+1};g];g]$, we have                
\bea
\ \ \ \ \ \ \hat S[\hat z[z^{k+1};g];g]=S_{0}[z^{k+1}]+\sum_{i=1}^{k+1}g^i
V_i
[\varphi^{k+1}]+g^{k+2} \tilde
S^{(k+2)}[z^{k+1}]+O(g^{k+3}),\label{332}
\eea
for some $\tilde S^{(k+2)}[z^{k+1}]$ with $s_0  V_i
=0$, which proves that if (\ref{ass}) holds,
the full Seiberg-Witten map can be obtained through an iteration
of canonical transformations:
\bea
z=\Pi_{k=1}^\infty\exp
g^k(\cdot,\Xi_{k})\hat z.
\eea
This discussion is in fact identical to the analysis
of the renormalization of non-power counting
renormalizable gauge theories as discussed in 
\cite{Voronov:1982ur,Weinberg:1996kr}.

Now, the crucial cohomological property
(\ref{ass}) that implies the existence of the
Seiberg-Witten map holds in the Yang-Mills case.  This is the
content of the cohomological theorems of 
\cite{Barnich:1994ve,Barnich:1995mt} proved
in the context of general deformations not limited by power-counting
restrictions or Lorentz-invariance. [As stated
above, there is an
antifield dependent deformation in the presence of a $U(1)$
factor (see also \cite{Brandt:2001hs}), 
as for $U(N)$ gauge groups; however, it
is easy to see that the noncommutative
deformation of Yang-Mills theory has no component
along this deformation because of its derivative structure,
see \cite{Barnich:2001bb}.] The fact that the 
Seiberg-Witten map for noncommutative $U(1)$
Yang-Mills theory can be extended to a canonical
transformation in field-anti\-field space follows from the
general properties relating canonical transformations
to redefinitions of the fields and of the gauge
transformations (see e.g. \cite{Henneaux:1992ig})
and has been explicitly verified in
\cite{Gomis:2000sp} (see also \cite{Brandt:2001aa}).
The cohomological property (\ref{ass}) also holds
for the Chern-Simons theory -- where in fact one
has the stronger result that $V_i$ is proportional to the
original action \cite{Barnich:2000zw} --,
providing a cohomological understanding of the results
of \cite{Grandi:2000av}.  In all these cases, there is
even a constructive (though somewhat cumbersome) procedure
for explicitly removing the antifield-dependent
terms and finding the generators of the successive
canonical transformations that bring the gauge structure
back to its original form through the use of
homotopy formulas.  A similar situation prevails for the
noncommutative Freed\-man-Town\-send model, as we shall
analyse in section \ref{4}.

Finally, it is also possible to
analyze along the above lines a situation where for instance the
deformation of the action and the gauge symmetries are non
trivial, while the algebra and higher order structure constants of
the gauge symmetries are equivalent.

\subsection{Weak Seiberg-Witten gauge equivalence \label{s3.4}}

By expanding the condition (\ref{3.35}) for the existence
of the Seiberg-Witten maps according to the antifield
number, one recovers formulas familiar from the
Yang-Mills context (but modified to weak relations).

For instance, if we denote by $f^i([\varphi],g)$ and
$g^\alpha([\varphi],[C],g)$ the expression of the
hatted fields and ghosts in terms of the original fields
and ghosts when the antifields are set equal to zero
in (\ref{transfo}),
\begin{equation}
f^i([\varphi],g)
= \frac{\delta^L F}{\delta \hat \varphi^*_i}
\big|_{\phi^* = 0}, \; \;
\; \;
 g^\alpha([\varphi],[C],g)
= \frac{\delta^L F}{\delta \hat C^*_\alpha}\big|_{\phi^* = 0}
\end{equation}
the condition (\ref{3.35}) becomes at antifield number zero
\bea
\hat I[f;g]=I^{\rm eff}[\varphi;g].
\label{eff}
\eea
In antifield number $1$, one gets
\bea
\hat \delta_{\hat\lambda} \hat \varphi^i\approx \delta_\lambda\hat
\varphi^i,\label{wesw}
\eea
where the even gauge parameter $\hat\lambda$ corresponds to the
odd function
$g^\alpha([\varphi],[C],g)$ while $\lambda$ corresponds to $C^\alpha$.
The relation \eqref{wesw} generalizes, in the open, reducible algebra
case, the key relation of \cite{Seiberg:1999vs}
that defines the Seiberg-Witten maps.

Finally, in antifield number $2$, 
one gets an integrability condition for
(\ref{wesw}), as well as possible (admissible)
 redefinition of the reducibility
functions. 
The integrability condition is the BRST version of the Wess-Zumino type
consistency condition \cite{Wess:1971yu} that one can deduce directly
from the
weak Seiberg-Witten equivalence condition (\ref{wesw}). 
In higher antifield number, one gets higher-order integrability
conditions related to the existence of higher-order structure
functions.

\mysection{Noncommutative Freed\-man-Town\-send model}

\subsection{Preliminaries}

We assume from now on
the space-time manifold to be ${\mathbb R}^n$ with
coordinates $x^\mu\,,\mu=1,\dots,n$. The Weyl-Moyal star-product is
defined through \bea f*g
(x)=\exp{i{\wedge_{12}}}
f(x_1)g(x_2)|_{x_1=x_2=x},\ \ \ \wedge_{12}=
\frac{g}{2}{\theta^{\mu\nu}}\partial_\mu^{x_1}\partial_\nu^{x_2}, \eea for a
real, constant, antisymmetric matrix $\theta^{\mu\nu}$. The
parameter $g$ is the deformation parameter. Standard
formulas are recovered by taking $g=1$, while the commutative
case corresponds to $g = 0$.

Let $M=m^A(x)T_A$ with $T_A$ generators of the Lie algebra $u(N)$,
i.e., antihermitian matrices. The coefficients $m^A(x)$ are real,
commuting or anticommuting fields.
If hermitian conjugation for the multiplication of ${\mathbb Z}_2$
graded functions is defined by $(mn)^\dagger=(-)^{|m||n|}
n m$, then hermitian conjugation of matrix valued
function also satisfies $(MN)^\dagger=(-)^{|m||n|} N
M$. 
We denote the invariant metric ${\rm  Tr}\ T_AT_B$
by $g_{AB}$, ${\rm Tr}\ T_AT_B=g_{AB}$.
It is non-degenerate.

The graded Moyal bracket defined by
\begin{equation}
[ M\stackrel{*}{,}
N]=M*N-(-)^{|M||N|}N* M
\end{equation}
is again a $u(N)$ valued function, because 
$(M*N)^\dagger = (-)^{|M||N|}N*M$.  This is a
straightforward extension of the reasoning of
\cite{Matsubara:2000gr} in the case where one allows the functions
to belong to a ${\mathbb Z}_2$ graded algebra. Furthermore, the
covariant derivative and associated field strength are defined as
follows: 
\bea 
\hat D_\mu M=\partial_\mu M+[\hat A_\mu\stackrel{*}{,}M],\cr
[\hat D_\mu,\hat D_\nu] M=[\hat F_{\mu\nu}\stackrel{*}{,} M],\cr
\hat F_{\mu\nu}
=
\partial_\mu \hat A_\nu-\partial_\nu\hat A_\mu
+ 
[\hat A_{\mu}\stackrel{*}{,}\hat A_\nu].
\eea
A key property of the Moyal star-product is
\bea
M* N = M N+\partial_\mu\Lambda^\mu. 
\eea 
As a
consequence, if boundary terms are neglected,
\bea
\int d^nx\ {\rm Tr}\ M*N=\int d^nx\ {\rm Tr}\  M N
=
\int d^nx\ {\rm Tr}\ N*M(-)^{|M||N|},\\
\int d^nx\ {\rm Tr}\  M* N* O
=
\int d^nx\ {\rm Tr}\ O* M* N(-)^{|O|(|M|+|N|)},\\
\int d^nx\ {\rm Tr}\  M*[N\stackrel{*}{,} O]
=
\int d^nx\ {\rm Tr}\ [M\stackrel{*}{,} N]* O,\\
\int d^nx\ {\rm Tr}\ \hat D_\mu  M* N
=
-\int d^nx\ {\rm Tr}\ M*\hat D_\mu  N. 
\eea

\subsection{Action and gauge algebra of noncommutative FT model}

The noncommutative $U(N)$ Freed\-man-Town\-send model exists
in four dimensions.  It is most conveniently
formulated in first order
form.  The action is  
\bea 
\label{eq:ncft-action}
\hat I=\int d^4x \ {\rm Tr}\
\Big(-\frac{1}{2} \epsilon^{\mu\nu\rho\sigma}\hat B_{\mu\nu}*\hat
F_{\rho\sigma}+\frac{1}{2}\hat A_\mu*\hat A^\mu\Big) 
\eea 
in Minkowski
space-time, with signature $(-+++)$, $\epsilon^{0123}=1$ and
$\epsilon_{0123}=-1$.                       
The action is
invariant under the gauge transformations \bea
\hat\delta_{\hat\lambda}\hat B_{\mu\nu}=\hat
D_{[\mu}\hat\lambda_{\nu]},\ \ \hat\delta_{\hat\lambda}\hat
A_{\nu}=0, \eea with gauge parameters $\hat\lambda_{\mu}$. The
gauge algebra is abelian, $[\hat\delta_{\hat\lambda^1},
\hat\delta_{\hat\lambda^2}]=0$ and reducible on-shell. Indeed, if
$\hat\lambda_\mu=\hat D_\mu\hat \eta$, 
then the gauge transformation reduces to on-shell trivial
transformations,
\bea \hat\delta_{\hat
D\hat\eta}\hat B_{\mu\nu}=\frac{1}{2}[\hat
F_{\mu\nu}\stackrel{*}{,}\hat\eta]=\frac{1}{4}\epsilon_{\mu\nu\rho\sigma}[
\frac{\delta\hat I}{\delta \hat
B_{\rho\sigma}}\stackrel{*}{,}\hat\eta]. \eea 

\subsection{Rigidity of gauge structure of commutative Freed\-man-Town\-send model}
\label{commft}    

If one sets $g = 0$, one gets the commutative Freed\-man-Town\-send
model, 
\bea  I_0=\int d^4x \ {\rm Tr}\
\Big(-\frac{1}{2} \epsilon^{\mu\nu\rho\sigma} B_{\mu\nu}
F_{\rho\sigma}+\frac{1}{2} A_\mu A^\mu\Big). 
\label{nncftm} 
\eea
with
\begin{equation}
F_{\mu\nu}=\partial_\mu 
A_\nu-\partial_\nu A_\mu+ [A_{\mu}{,}
A_\nu].
\end{equation}
This model is invariant under the ordinary gauge
transformations
\begin{equation}
\delta_\lambda B_{\mu\nu}= D_{[\mu}\lambda_{\nu]},\ \ \
\delta_\lambda A_{\nu}=0,
\label{gtcftm}
\end{equation}
with on-shell gauge reducibility
for $\lambda_\mu=  D_\mu\eta$.
The action (\ref{nncftm}) is the sum of a free, quadratic
part plus a cubic interaction vertex proportional to
\begin{equation}
g_{AD}f^D_{\; \; \; BC} \epsilon^{\mu\nu\rho\sigma} B^A_{\mu\nu} 
A^B_\rho A^C_\sigma
\label{cubicvertex}
\end{equation}
where $f^A_{\; \; \; BC}$ are the structure constants of
$u(N)$.

Obvious consistent deformations of the commutative
Freed\-man-Town\-send model are obtained by adding to
(\ref{nncftm}) an arbitrary polynomial in the $A_\mu$'s
and their derivatives.

Because these are strictly gauge invariant under (\ref{gtcftm})
they do not modify the gauge structure.  It turns out
that these are the most general consistent deformations.
Indeed, it has been shown in \cite{Henneaux:1996ts} that the
Freed\-man-Town\-send vertex (\ref{cubicvertex}),
characterized by general structure
constants $C^A_{\; \; \; BC}$ of a Lie-algebra,
is the only gauge-symmetry deforming interaction vertices
for a set of abelian $2$-forms in $4$ dimensions.  That is,
the most general deformation of (\ref{nncftm}) with $f^A_{\; \; \; BC}=0$
is (\ref{nncftm}) with $f^A_{\; \; \; BC}$ replaced by
the structure constants
of an arbitrary Lie algebra, plus terms that involve only
$A_\mu$ and its derivatives
(see also
\cite{Henneaux:1997ha,Bizdadea:2000hb}).  [There is no
possibility for a Chern-Simons term $H \wedge B$, where $H$ is
the field strength of $B$, because $H \wedge B$ is a $5$-form.]
It easily follows from that result
that the general first order deformation of the $u(N)$-Freed\-man-Town\-send
model (\ref{nncftm}) is given again by the Freed\-man-Town\-send action but
with $f^A_{\; \; \; BC}$ replaced by $f^A_{\; \; \; BC} +
g m^A_{\; \; \; BC}$, plus terms that involve only
$A_\mu$ and its derivatives.  The constants $m^A_{\; \; \; BC}$
are constrained by $m^A_{\; \; \; B[C} f^B_{\; \; DE]} +
f^A_{\; \; \; B[C} m^B_{\; \; DE]}$ (Jacobi identity), i.e.,
define first-order deformations of the Lie algebra $u(N)$.
But this algebra is rigid (there is only one abelian
factor) in the sense that one can bring
$f^A_{\; \; \; BC} +
g m^A_{\; \; \; BC}$ back to $f^A_{\; \; \; BC}$ by linear
redefinitions in internal space. These are just particular
canonical transformations (they would generate
terms quadratic in $A^A_\mu$ if the metric is not
invariant, but this is a deformation of the
announced type).  Thus, we may actually assume
$m^A_{\; \; \; BC} = 0$, which means that up to redefinitions,
the only deformations of the Freed\-man-Town\-send
model are precisely exhausted by the addition of 
functions of
the $A_\mu$ and their derivatives.    
This implies, in particular, that the noncommutative
Freed\-man-Town\-send model must be amenable to the form
\bea
\hat I=I_0+V[A;g].
\eea

Note that the field $A_\mu$ is auxiliary in (\ref{nncftm})
-- i.e., can be eliminated through its own equation
of motion.  It remains auxiliary in the deformed theory
in the sense that its equations of motion can 
still be solved as formal
power series in $g$.  Note also that the equations of
motion for $B_{\mu\nu}$ are left unchanged in the
deformation and imply $F_{\mu\nu} = 0$.  
By solving
this constraint, $A = g^{-1} dg$, the action becomes that
of the non-linear sigma model modified by higher dimensionality
operators.  

\mysection{Seiberg-Witten map for Freed\-man-Town\-send model
\label{4}}

\subsection{Existence of the Seiberg-Witten map}
\label{s41} 

The solution of the master equation for the noncommutative
Freed\-man-Town\-send model is given by
\begin{multline}
\hat S=\int d^4x \ {\tr}\ \Big( -\frac{1}{2}
\epsilon^{\mu\nu\rho\sigma}{\hat B}_{\mu\nu}*\hat F_{\rho\sigma}
+\frac{1}{2}\hat A_\mu*\hat A^\mu+\\
+ {\hat B}^{*\mu\nu}*\hat D_{[\mu} \hat C_{\nu]} + {\hat
C}^{*\mu}*{\hat D}_\mu {\hat \rho} + \frac{1}{8} \qcommut{\hat
B^{*\mu\nu}}{{\hat B}^{*\rho\sigma}} \epsilon_{\mu\nu\rho\sigma} *
{\hat \rho} \Big)\,. \label{eq:FT-master}
\end{multline}      
while the solution of the master equation for the
commutative case is given by the same expression with the
$*$-product replaced by the ordinary product \cite{Batlle:1988rd},
\begin{multline}
S_0=\int d^4x \ {\tr}\ \Big( -\frac{1}{2}
\epsilon^{\mu\nu\rho\sigma}{\hat B}_{\mu\nu}\hat F_{\rho\sigma}
+\frac{1}{2}\hat A_\mu\hat A^\mu+\\
+ {\hat B}^{*\mu\nu}\hat D_{[\mu} \hat C_{\nu]} + {\hat
C}^{*\mu}{\hat D}_\mu {\hat \rho} + \frac{1}{8} \commut{\hat
B^{*\mu\nu}}{{\hat B}^{*\rho\sigma}} \epsilon_{\mu\nu\rho\sigma} 
{\hat \rho} \Big)\,. \label{eq:FTcc-master}
\end{multline}           

It follows from the cohomological analysis of the previous section
that the Seiberg-Witten maps exists: by a canonical transformation,
one can transform the functional (\ref{eq:FT-master}) into
the solution of the master equation for the commutative
theory, plus $V_i([A_\mu])$-terms that are strictly
gauge-invariant\footnote{Another way to
arrive at the same conclusion, valid for Lie algebras that
are non-rigid,  is the following. The only obstruction to
the SW map arises if one ``hits"
the cubic vertex (\ref{cubicvertex}) in
the noncommutative deformation process.  But this cannot be the
case, because of a direct power counting argument: the
vertex (\ref{cubicvertex}) has dimension $5$, while the first
noncommutative vertex has dimension $6$ -- the field $B$ has dimension
$1$ while the auxiliary field $A$ has dimension $2$.
}.                                               

One can explicitly construct the canonical transformation
order by order in the fields, using standard cohomological 
weapons (homotopy formula for the free BRST differential etc).
We have not been able to sum the formal series obtained in
this recursive manner in a concise and
useful  way, however, except in the $u(1)$-case, to which we shall
therefore now exclusively turn.

\subsection{SW map in the
$U(1)$ case}
In the $u(1)$-case, the solution of the master
equation for the commutative model is given by
\begin{equation}
 S_{0}[\phi,\phi^*]=\int d^4x
\Big( -\frac{1}{2} \epsilon^{\mu\nu\rho\sigma}{
B}_{\mu\nu}F_{\rho\sigma} +\frac{1}{2} A_\mu A^\mu +{ B}^{*\mu\nu}
\d_{[\mu}  C_{\nu]} +{ C}^{*\mu} {\d}_\mu { \rho}\Big)\,,
\label{eq:FTA-master-commut}
\end{equation}                
Our goal is to find a canonical transformation
(\ref{transfo}) such
that (\ref{3.38}) is satisfied for $\hat S[\hat\phi,\hat\phi^*]$ given
by \eqref{eq:FT-master} and $S_0[\phi,\phi^*]$ by
\eqref{eq:FTA-master-commut}.
The searched-for
generating functional $F[\phi,\hat\phi^*]$ 
takes the form
\begin{multline}
\label{eq:U1-F}
  F[\phi,\hat\phi^*]= \int d^nx\ \Big( {\hat A}^{*\mu}f_\mu([A],[H])
  +\hat B^{*[\mu\nu]}f_{[\mu\nu]}([A],[B]) + {\hat
    C}^{*\mu}(C_\mu+2\qcommut{C_{\alpha}}{f_\mu}^\alpha)\\
  - \frac{1}{4} \epsilon_{\mu\nu\rho\sigma} {\hat B}^{*\mu\nu}
  \qcommut{C_\alpha}{{\hat B}^{*\rho\sigma}}^\alpha +\hat\rho^*\hat
  \rho \Big)\,.
\end{multline}
where the bracket $[\cdot\stackrel{*}{,}\cdot]^\alpha$ is defined
by \bea
\qcommut{f}{h}^\mu = -i\frac{g}{2}\theta^{\mu\nu} f*_1\d_\nu h
,\label{A66}\eea with \bea
f*_1g(x)=\frac{\sin{\wedge_{12}}}{{\wedge_{12}}}
f(x_1)g(x_2)|_{x_1=x_2=x} ,\eea
(see \cite{Mehen:2000vs} and \cite{Liu:2000mj} for further information
on $*_1$).
By construction, this bracket
satisfies
\bea
[\partial_\mu f
\stackrel{*}{,}h]^{\mu}=-\half\qcommut{f}{h}.
\eea
It is easy to check that in antifield nubmber higher than 1,
the generating functional~\eqref{eq:U1-F} satisfies~\eqref{3.38}
for arbitrary $f_\mu([A],[H])$.

Identifying terms of antifield number 1
in \eqref{3.38} one gets,
\begin{equation}
f_{\mu\nu}=B_{\mu\nu}+2
\qcommut{B_{\mu\alpha}}{f_\nu}^\alpha -2
\qcommut{B_{\nu\alpha}}{f_\mu}^\alpha
-\frac{ig\theta^{\alpha\beta}}{4} B_{\alpha\beta}
*_1 {\hat F}^f_{\mu\nu}+[B_{\alpha\beta}\stackrel{*}{,}
f_\mu\stackrel{*}{,}f_\nu]^{\alpha\beta},
\end{equation}
where
\begin{equation}
\hat F^f_{\mu\nu}=\d_\mu f_\nu -\d_\nu f_\mu + \qcommut{f_\mu}{f_\nu}
\end{equation}
and
\begin{equation}
[B_{\alpha\beta}\stackrel{*}{,}
f_\mu\stackrel{*}{,}f_\nu]^{\alpha\beta}
=
\sigma_0 \Big(\qcommut{f_{\mu}}{\qcommut{C_{\alpha}}{f_\nu}^\alpha}-
\qcommut{f_{\nu}}{\qcommut{C_{\alpha}}{f_\mu}^\alpha}-
\qcommut{C_\alpha}{\qcommut{f_\mu}{f_\nu}}^\alpha \Big)\,.
\end{equation}
Here, $\sigma_0$ is a particular contracting homotopy for the differential
$\gamma_0$ (longitudinal exterior derivative of the free theory)
given in \cite{Henneaux:1998rp}). Explicitly, if instead of
the variables $B_{\mu\nu}$, $C_\mu$, $\rho$, and their derivatives,
we define new variables through 
\bea
y^\alpha\equiv
\partial_{(\nu_1}\dots
\partial_{\nu_{l}} B_{{[\mu)}\lambda]}
,\partial_{(\nu_1}\dots\partial_{\nu_l} C_{\mu)},\nonumber\\
z^\alpha \equiv
\partial_{(\nu_1}\dots
\partial_{\nu_{l}}\partial_{[\mu)}C_{\lambda]},
-\partial_{(\nu_1}\dots\partial_{\nu_l} \partial_{\mu)} \hat \rho,\\
\rho,
\partial_{(\nu_1}\dots
\partial_{\nu_{l}} H_{\mu)\rho\sigma},
\eea
for $l=0,1,\dots$ and
$H_{\mu\rho\sigma}=\partial_{[\mu}B_{\rho\sigma]}$, 
a particular expression for the contracting homotopy $\sigma_0$ 
is given by 
\bea
\sigma_0 f =\int^1_0 \frac{dt}{t} [y^\alpha\frac{\partial f}{\partial
  z^\alpha}](ty,tz). 
\eea

We leave it to the reader to check
that the generating functional $F$ does
the job of transforming \eqref{eq:FT-master} 
into \eqref{eq:FTA-master-commut} no matter how the gauge-invariant functions
$f_\mu([A],[H])$ is chosen (it must just be invertible). That there
is an ambiguity in the SW map, characterized
by addition of the gauge-invariant
functions to $f_\mu$, $f_{\mu\nu}$ and
also to the higher antifield number terms in $F$
(so that there are in fact {\em maps})
is not surprising: any
redefinition that involves only gauge invariant quantities preserves
the gauge structure.

To summarize, a particular class of solutions for the SW map for
the noncommutative abelian Freed\-man-Town\-send model has been obtained.
The new feature of this map, compared with the SW map for Yang-Mills
models, is that the generating functional $F[\phi,\hat\phi^*]$ is
quadratic in some of the antifields, so that the transformations of
some of the fields contain the antifields.
This is related to
the fact that the equations of motion appear in the
reducibility identities (while the Yang-Mills
gauge structure is defined not just on-shell, but
also off-shell).  This feature
is easily incorporated at no cost
in the antifield formalism.  

Finally, we note that the Lagrangian of the non-commutative
Freed\-man-Town\-send model can be mapped to the Lagrangian of
the commutative one up to second order in $\theta$.  It is an
interesting
question to investigate whether this holds to all orders.  In
this context, we recall that the 2-dimensional commutative and
noncommutative
WZW models are known to be equivalent not just in their gauge
structure but also in their action
\cite{Nunez:2000rs,Moreno:2000kt}.

\section{Conclusion\label{5}}

The conclusion in \cite{Madore:2000en} is ``that there should be
an underlying geometric reason for the Seiberg-Witten map.'' The
analysis of this paper shows that there is at least a deformation
theoretic reason for the existence of this map in the following
sense.

Consistent deformations of gauge theories with non trivial
deformations of the gauge structure are in general severely
constrained. 
The appropriate framework to study these constraints
in the general case (reducibility, closure only on-shell)
is the an\-ti\-field-an\-ti\-brac\-ket formalism.   The
cohomology that controls non trivial deformations of the gauge
structure is the local BRST cohomology, and how
non trivial cocycles depend on the antifields.

By analyzing explicitly the noncommutative Freed\-man-Town\-send
model, these considerations have been shown to apply beyond the
original Yang-Mills framework. Because they do not depend on the
precise deformation considered, they also apply for deformations
that involve for instance more complicated star-products than the
Weyl-Moyal star-product.  The analysis can also be straightforwardly
extended to models with higher rank $p$-forms as
discussed in \cite{Henneaux:1997ha}.

The cohomological theorems of \cite{Barnich:1994ve,Barnich:1995mt,
Barnich:2000zw}, which guarantee the existence
of the SW maps were studied initially with quantum motivations
in mind (they control perturbative renormalizability 
-- i.e., gauge invariance of the needed counterterms -- as
well as candidates
anomalies for general (effective) theories with 
the same gauge structure as Yang-Mills
models).  The present paper clearly indicates their 
relevance in the classical context as well.

\section*{Acknowledgments}
M. H. is grateful to Kostas Skenderis for informative discussions
at an early stage of this work.
This research has been partially supported by the ``Actions de
Recherche Concert{\'e}es" of the ``Direction de la Recherche
Scientifique - Communaut{\'e} Fran{\c c}aise de Belgique", by
IISN - Belgium (convention 4.4505.86),  by
Proyectos FONDECYT 1970151 and 7960001 (Chile), 
by the European Commission RTN programme HPRN-CT-00131,
in which the authors are associated to K. U. Leuven and 
by INTAS grant 00-00262. The work of M. G. is partially supported by 
the RFBR grant 99-01-00980. He is grateful to the Free University of
Brussels for kind hospitality.

\newpage

\providecommand{\href}[2]{#2}\begingroup\raggedright\endgroup

\end{document}